\documentclass[10pt,onecolumn]{article}
\usepackage{times}
\usepackage{graphicx}
\usepackage{amssymb}
\usepackage{url,hyperref}
\usepackage{upgreek}
\begin{document}
\title{Fast 3D cell tracking with wide-field fluorescence microscopy through deep learning}

\author{Kan Liu$^{2,\dag}$, Hui Qiao$^{1,\dag}$, Jiamin Wu$^1$, \\
Haoqian Wang$^3$, Lu Fang$^2$, Qionghai Dai$^{1,2,*}$\\
\\
$^1$ Department of Automation, Tsinghua University, China\\
$^2$ Tsinghua-Berkeley Shenzhen Institute, Tsinghua University, China\\
$^3$ Graduate School at Shenzhen, Tsinghua University, China\\
\\
$^*$ Corresponding author: qhdai@tsinghua.edu.cn
}






\maketitle
\thispagestyle{empty}

\begin{abstract}
Tracking cells in 3D at high speed continues to attract extensive attention for many biomedical applications, such as monitoring immune cell migration and observing tumor metastasis in flowing blood vessels. Here, we propose a deep convolutional neural networks (CNNs) based method to retrieve the 3D locations of the fluorophores from a single 2D image captured by a conventional wide-field fluorescence microscope without any hardware modification. The reported method converts the challenging 3D localization from an ill-posed model-based fitting problem, especially with dense samples and low signal-to-noise ratio, to a solvable multi-label classification problem through two cascaded CNNs, where deep learning technique has a great advantage over other algorithms. Compared with traditional kernel fitting methods, the proposed method achieves more accurate and robust localization of multiple objects across a much larger axial range, which is validated by both simulation and experimental results on 3D distributed fluorescent beads. Moreover, \emph{in vivo} 3D tracking of multiple blood cells in zebrafish at 100 fps further verifies the feasibility of our framework.
\end{abstract}



\maketitle

\section{Introduction}

Understanding the dynamic mechanisms and cell-cell interactions that underlie the biological processes of complex systems, such as immune cell migration \cite{germain2006dynamic} and tumor metastasis \cite{steeg2006tumor}, requires imaging methods that are capable of tracking living cells simultaneously across large volumes of tissue. Moreover, tracking gene-expressed cell is an essential way to monitor the therapeutic response, which is one of the most critical issues for ensuring success of gene therapy \cite{shah2004molecular}. For \emph{in vivo} imaging, inevitable motion blur induced by the respiration or heart beating makes it more challenging to track cells consecutively.

Over the past decades, three-dimensional (3D) localization microscopy has been rapidly evolved \cite{huang2008three,sauer2013localization,legant2016high} and drawn increasing interest in biomedical research. A variety of imaging modalities have been developed to meet the modern requirement. Scanning imaging modalities, including two-photon microscopy \cite{denk1990two,song2017volumetric}, light-sheet microscopy \cite{keller2008reconstruction}, and confocal fluorescence microscopy \cite{jeong2016high}, always need specific hardware modifications and successively scan the 3D volume of the biological sample. Although the good optical sectioning ability provides a more accurate localization in axial dimension with a better signal-to-noise ratio (SNR), the scanning process decreases the time resolution, restricting its application for capturing dynamic events or moving specimens in a big volume. Light-field microscopy (LFM) \cite{levoy2006light} measures both the spatial and angular distribution of the light propagating through the sample volume in a single shot. The 3D deconvolution \cite{prevedel2014simultaneous} or the signature-fitting algorithm \cite{pegard2016compressive} can be incorporated into LFM for high-speed volumetric imaging, but the setup still needs elaborate optical design and calibration.

In contrast to LFM, wide-field fluorescence microscopy (WFM), using a conventional inverted microscope without any hardware modification, is the most commonly accessible tool of every biological and biomedical laboratory. WFM detects all the in-focus and out-of-focus fluorescent probes simultaneously without scanning. The out-of-focus photons are usually regarded as the background noise which people try to get rid of. Given that the measured pixelated shape of each defocus spot still contains the valid information, kernel fitting methods \cite{shechtman2016multicolour,wang2017single} are adopted to localize the 3D position of the defocus spot for further deconvolution. Unfortunately, such kernel fitting methods require a priori knowledge of the 3D point spread function (PSF) and most of them fail for densely labeled specimens. Besides, the fitting process is ill-posed and always results in an uncontrolled noise amplification \cite{tikhonov1977methods}. Recently, deep learning \cite{lecun2015deep}, widely adopted by the computer-vision community, is making major advances in image classification \cite{ren2015faster}, image super-resolution \cite{dong2016image}, image deconvolution \cite{xu2014deep} and many other studies. Despite the convolutional neural networks (CNNs) have been successfully applied to optical imaging \cite{sinha2017lensless,rivenson2017deep}, more potential of such an end-to-end learning framework remains to be exploited in the biomedical optics field.

Here, we propose a deep convolutional neural networks based method for fast 3D cell tracking with normal wide-field fluorescence microscopy. Concretely, two cascaded CNNs, named as lateral detection CNN and axial localization CNN respectively, are constructed to obtain 3D localization with a single snapshot, making it possible to perform fast 3D tracking at the frame rate of the camera sensor. Rather than building a complicated model with numbers of sensitive parameters to handle diverse imaging conditions, such as optical aberrations and different noise statistics, we convert the ill-posed fitting problem to a cascaded classification problem. A large set of training data can be synthesized by a simple calibration step for the deep CNNs to deal with the various situations that we may face. Compared with previous kernel fitting methods, our method achieves a more precise and robust 3D localization. We also demonstrate its performance experimentally for tracking zebrafish blood cells close to the heart at $100$ fps. Since the method is general under different optical systems and arbitrary illumination patterns, our end-to-end training framework can be easily extended to other PSF engineering methods. Although the scope of this paper lies in the 3D localization microscopy, we believe our work bridges the gap between micro and macro areas and will further promote both the optical and computer vision communities. Our approach can also open rich perspectives for the potential of deep neural networks in the optical society.

\section{Methods}

\subsection{Experimental Setup}

Image data acquisition was performed using a Zeiss Axio Observer.Z1 wide-field microscope equipped with a motorized stage and controlled by Micro-Manger software. The microscope used a mercury lamp for illumination and an infinity color corrected objective lens (Zeiss Plan-Apochromat, $10\times/0.45$) with a numerical aperture (NA) of $0.45$. The fluorescence images were obtained by a digital sCMOS camera (Hamamatsu ORCA-Flash4.0 V2, C11440-22CU) with a pixel size of $6.5$ $\upmu$m.

\subsection{Deep Learning Based Method}

\begin{figure*}[t]
\centering
\includegraphics[width=\linewidth]{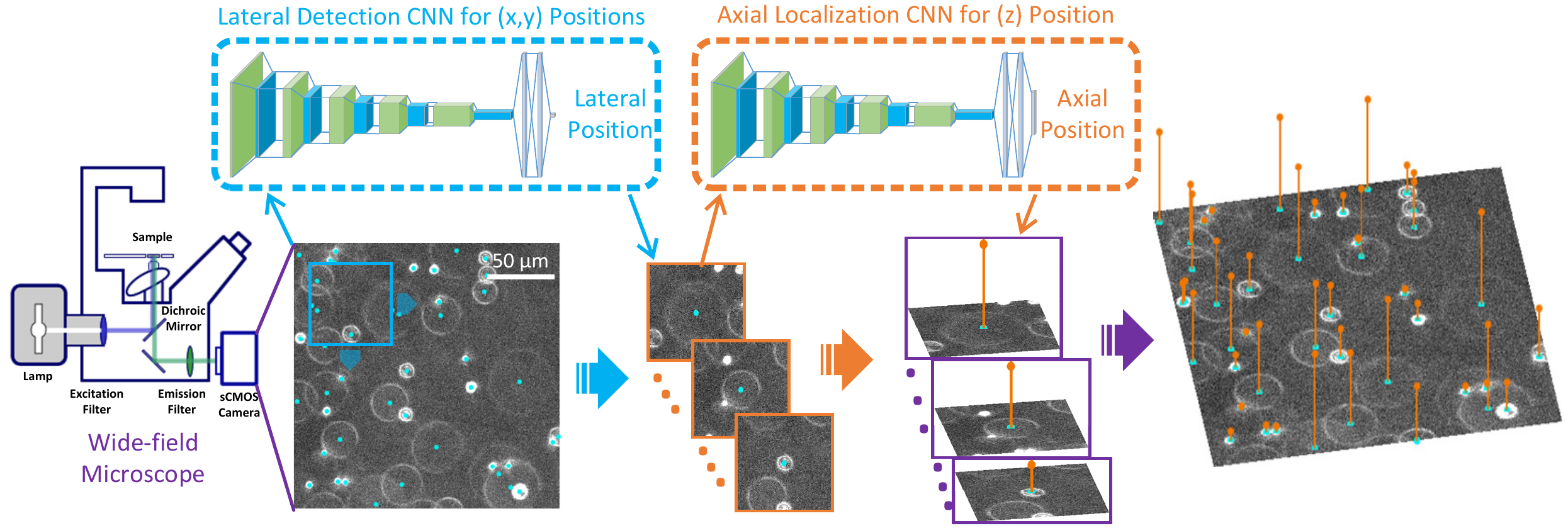}
\caption{Framework of the proposed 3D localization microscopy. Lateral detection CNN, highlighted by the blue dashed box, first determines whether there exist diffraction patterns at the central lateral position of the sliding window. Axial localization CNN, highlighted by the orange dashed box, then estimates the axial positions of the predicted positive samples of lateral detection CNN.}
\label{fig:framework}
\end{figure*}

As shown in Fig.~\ref{fig:framework}, the framework of our proposed 3D localization microscopy consists of two cascaded convolutional neural networks, denoted as lateral detection CNN and axial localization CNN, respectively. On account of the assumption that the PSF of a normal microscope is translation-invariant at each depth, we use a sliding window and decompose the 3D localization into lateral and axial phases for dimensionality reduction and parallel computation. The size of sliding window is dependent on the lateral size of PSF and is not necessary to be too large due to the low power density at large defocus distance. After capturing a wide-field image, lateral detection CNN first predicts whether there exist diffraction patterns at the central $x$-$y$ (lateral) position of the concerned sliding window. Consequently, we can detect the fluorescent probes and estimate their lateral positions. At the next step, axial localization CNN identifies the axial position of each detected fluorescent probe according to its corresponding diffraction pattern. Therefore, 3D positions of the fluorescent probes are finally acquired. Making use of the determined 3D localization results, fast 3D tracking can be realized with a Kalman filter \cite{julier1997new}. We construct these two CNNs in the light of the architecture of VGG-19 network \cite{simonyan2014very}, where lateral detection CNN is built as a binary classifier while axial localization CNN acts as a multi-label classifier in order to simultaneously recognize the different axial positions of multiple diffraction patterns that have the same lateral position. Both architectures are composed of several functional layers: convolution layer, pooling layer, fully connected layer, and output layer (see Section 1 of supplementary material for detailed architectures).

To successfully train the deep CNNs, we need a substantial collection of labeled training data. Using a $1$-$\upmu$m fluorescent bead as a sub-resolution point, we initially recorded a single focal stack by moving the stage from the focal plane to $100$ microns above with $2$-micron spacing. Based on the captured focal stack, we can simulate vast quantities of wide-field images via sampling spatial locations of fluorescent beads independently and randomly from a 3D space. We then altered the density of beads and fluorescence intensities to further augment the dataset. Different scales of Poisson noise were also added to enhance the robustness of our method when encountering the real data. We finally extracted $128\times128$-pixel (sliding window size) image patches and selected $200000$ examples with their assigned labels (see Fig.~2 in supplementary material) for each network training.

We implemented the CNNs using MatConvNet \cite{Vedaldi2015MatConvNetCN}, and the training process was running on a Windows 8 environment with CPU Intel i7-6850k, 2 Nvidia GTX 1080 Ti graphic cards and 128 GB memory. In this training phase, we set the learning rate empirically to 0.001, and allowed it to decay exponentially with a rate of 0.0005 as the training progresses. A dropout technique was applied on the first fully connected layer in training with a probability of 0.5, while in validation and testing it was set to be 1, i.e., no dropout. This allows $50\%$ of the nodes to be randomly chosen and intentionally disabled in training to prevent over-fitting, while keeping all nodes alive can effectively check how well the network has learned \cite{ren2018learning}. Every time only a small batch of 8, called a mini-batch, of the entire training data was fed into the network for training, so that we were able to tackle the issues of limited computer memory and stagnation in local minima during optimization. For random initialization, we sampled the weights from a normal distribution with the 0 mean and $10^{-1}$ standard deviation. The biases were initialized with 0. We utilized the stochastic gradient descent with momentum as the optimization algorithm. In each mini-batch training, we performed one iteration of the optimization and then updated the parameters of the network. The whole training stage was stopped after $5$ epochs. The processing time to run the learning algorithm (5 epochs with the batch size of 8) for each network was nearly 7 hours.

After the training procedure, which needs to be performed only once, the CNNs are fixed and ready to accomplish 3D localization blindly. To measure the localization performance, we first match each localized point to a nearby ground truth (GT) point which has the minimal Euclidean distance from it. Localized point, with a distance closer than a tolerance radius ($10$ $\upmu$m in our experiments) from its match, is then considered as true positive (TP). The remaining localized points are categorized as false positives (FPs). Missed GT points are called false negatives (FNs). We compute the Jaccard index J as follows:
\begin{equation}
{\rm J} = \frac{{\rm TP}}{{\rm FN} + {\rm FP} + {\rm TP}},
\label{eq:eq1}
\end{equation}
and a Jaccard index of $1.0$ indicates a perfect localization performance. Additionally, we introduce another metric: the root-mean-squared error (RMSE) between the 3D positions of TPs and their matched GT points.

\section{Results}

\subsection{Spatial Resolution}

\begin{figure}[t]
\centering
\includegraphics[width=0.75\linewidth]{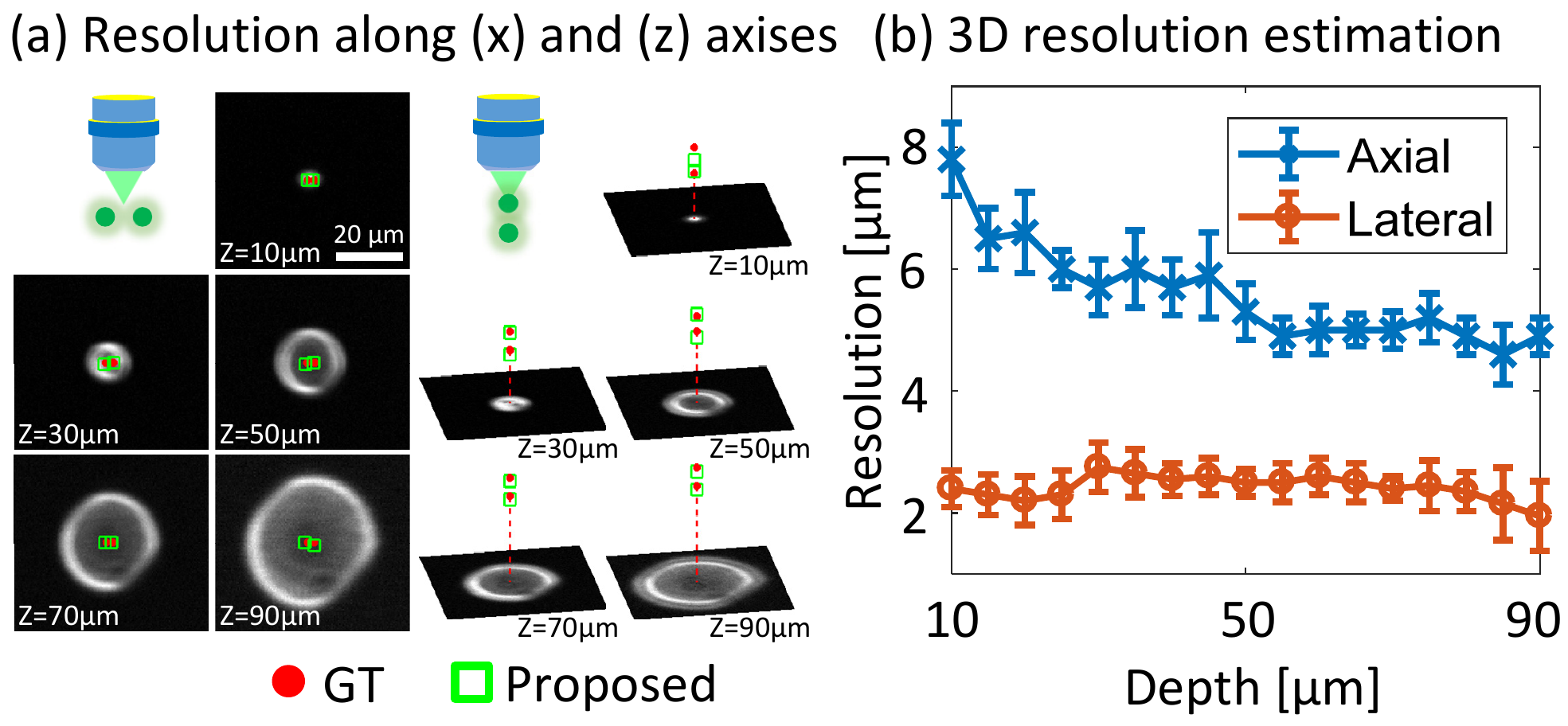}
\caption{Spatial resolution analysis for our method. (a) Resolvable examples at different depths, according to the minimal distance between two points required for correct identification as separate points in lateral dimension and axial dimension. (b) Estimated lateral resolution and axial resolution as a function of depth.}
\label{fig:resolution}
\end{figure}

To quantitatively analyze the localization resolution of our method with multiple objects, we conducted a simulation experiment to find out the required minimal distance for the separation of two points (with the same intensities) positioned in lateral dimension and axial dimension, respectively. The point spread function used for the resolution analysis was captured with a $1$-$\upmu$m fluorescent bead by the epifluorescence microscope described before. The details of the simulation are illustrated in Fig.~\ref{fig:resolution}. We gradually reduced the distance between the two points (labeled by red spots) and used our algorithm to localize (labeled by green squares). When our method cannot distinguish these two points and recognized them as a single point at a certain distance, we marked this minimum distance as the resolution.

Both lateral case and axial case are demonstrated at different depths ranging from $10$ $\upmu$m to $90$ $\upmu$m, respectively. Representative examples are also shown in the Fig.~\ref{fig:resolution}(a). Fig.~\ref{fig:resolution}(b) illustrates the result of the resolution evaluation, with $\sim$2 $\upmu$m in lateral dimension and $\sim$6 $\upmu$m in axial dimension, which are quite close to the pixel-size-limited system resolution of $1.3$ $\upmu$m in lateral dimension and $5$ $\upmu$m in axial dimension. More specifically, the lateral resolution is rather uniform across different depths, while the axial resolution tends to decrease when approaching the focal plane due to the difficulty of demixing the overlapped tiny diffraction patterns. Although it may pose a practical limitation for our method, a combination with the PSF engineering technique can alleviate this problem effectively.

\subsection{3D Localization Performance Analysis on Simulated Data}

Conventional localization microscopy with various kernel fitting methods \cite{shechtman2016multicolour,balzarotti2016nanometer} has an intrinsic limitation in localization accuracy caused by the SNR of the measurement. And the simultaneous localization of multiple objects further increases the localization error due to the intensity crosstalk of different sources. By including the complicated distributions and noise conditions into the training data, our method automatically optimizes the features for better and more robust localization. To validate the performance enhancement of our method, compared with the maximum likelihood estimation (MLE) algorithm \cite{shechtman2016multicolour} used for the state-of-the-art kernel fitting methods, we demonstrated two simulation experiments comparing the two approaches under different SNRs and different sample densities, respectively.

\begin{figure}[t]
\centering
\includegraphics[width=0.75\linewidth]{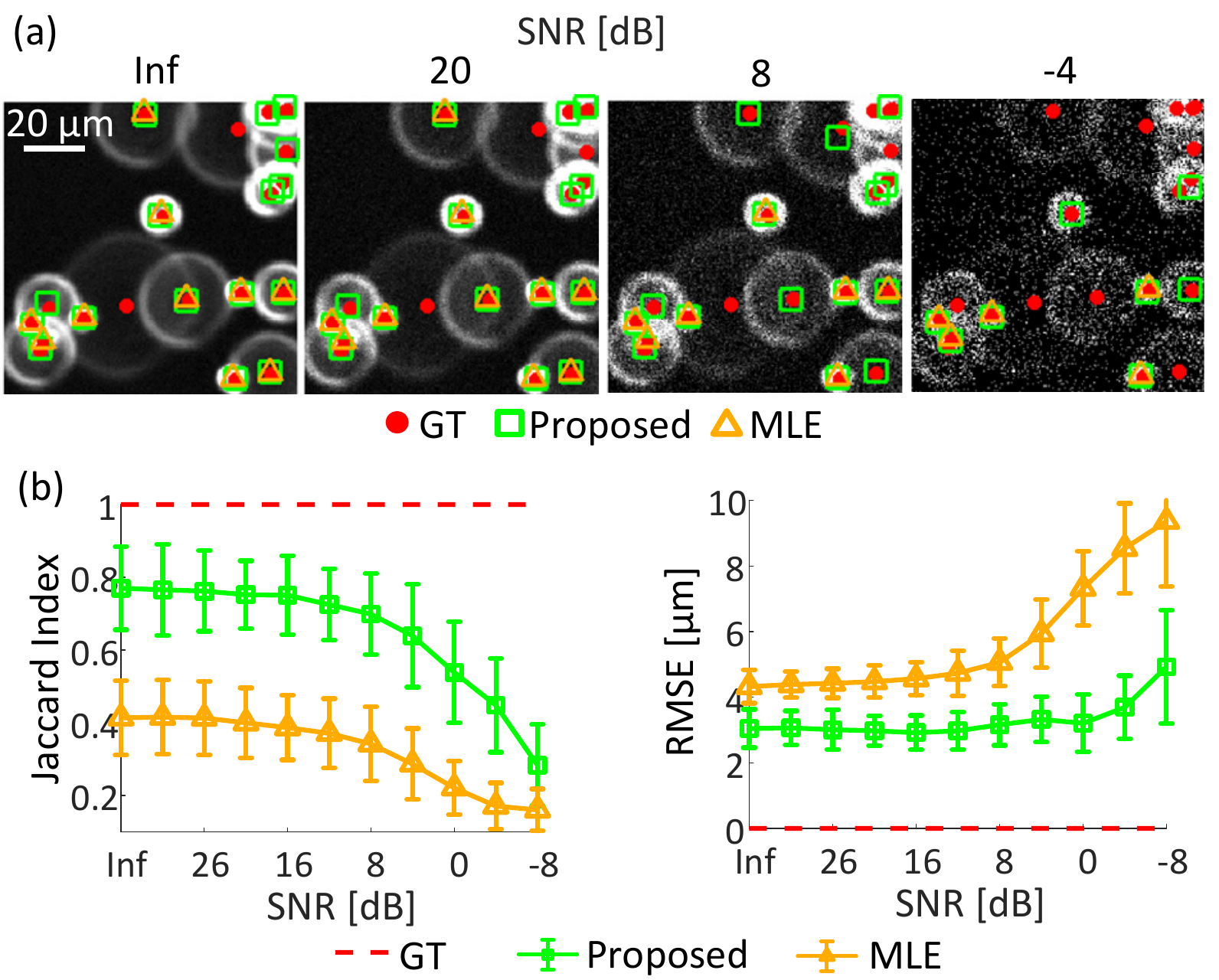}
\caption{Localization with different SNRs. (a) Comparison of the localization results under different SNR conditions reconstructed by our method and MLE method. (b) The curves of Jaccard index and RMSE versus the SNR, respectively.}
\label{fig:snr}
\end{figure}

Firstly, $20$ points were randomly placed in the volume of $100\times100\times100$ $\upmu$m$^3$. Intensities of the points were randomly sampled from 0.1 to 1 and a wide-field fluorescence image was then simulated with the previous captured PSF. Different levels of Poisson noise were added to the simulated image, corresponding to different SNR conditions. Then we applied our method and MLE method to localize these points, respectively, as shown in Fig.~\ref{fig:snr}(a). The Jaccard index and RMSE of each reconstruction are plotted as the curves versus different SNRs in Fig.~\ref{fig:snr}(b). Our method achieves a much higher recognition accuracy compared with MLE under different SNR conditions. Even when the SNR decreases to $0$ dB, our method still has a Jaccard index above $0.5$, which is even higher than the Jaccard index of MLE method without noise. Examining the RMSE curve, MLE method shows an increasing localization error with reduced SNR, which is the inherent limitation of kernel fitting methods. On the contrary, our method works more robustly, inducing low and consistent RMSE against different noise levels.

\begin{figure}[t]
\centering
\includegraphics[width=.75\linewidth]{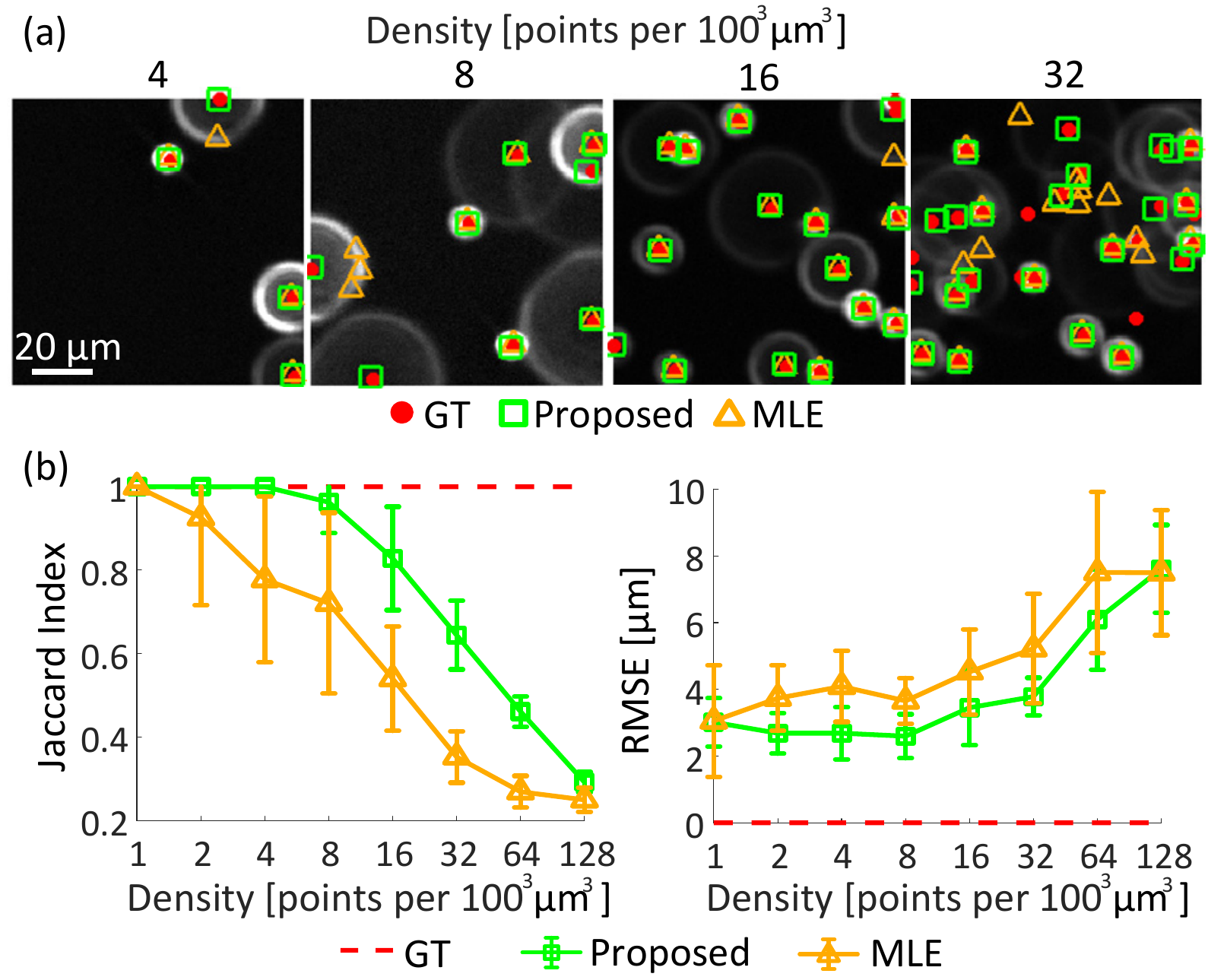}
\caption{Localization with different sample densities. (a) Comparison of the localization results under different sample densities reconstructed by our method and MLE method. (b) The curves of the Jaccard index and RMSE versus the density, respectively.}
\label{fig:density}
\end{figure}

To demonstrate the 3D localization ability of our method with multiple objects in a single image, we randomly placed $1$ to $100$ points in a $100\times100\times100$ $\upmu$m$^3$ volume with the same simulation process as before. The localization results with 4, 8, 16, and 32 points are shown in Fig.~\ref{fig:density}(a). Due to the crosstalk of high-density objects, MLE method almost fails and has a much lower Jaccard index compared with our method, as shown in Fig.~\ref{fig:density}(b). Particularly, the localization errors of both methods become larger with the increase of the sample density, while our method has a smaller slope and higher accuracy.

\subsection{3D Localization of Multiple Fluorescent Beads}

\begin{figure}[t]
\centering
\includegraphics[width=.75\linewidth]{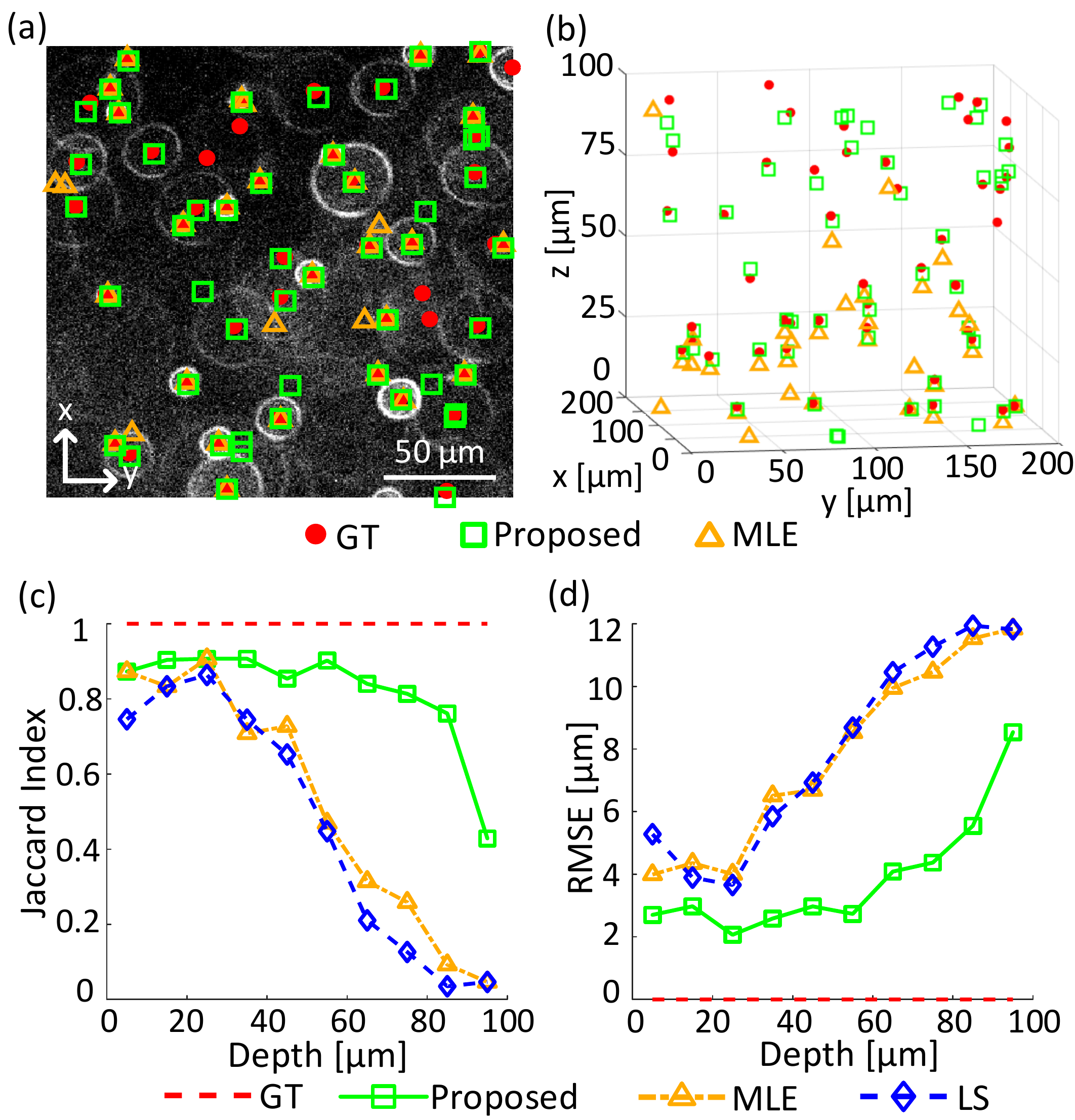}
\caption{Single-shot wide-field fluorescence image and 3D localization of randomly placed fluorescent beads. (a) and (b) Collected wide-field fluorescence image and comparison of the localization results reconstructed by our method and MLE method. (a) Estimated source positions are projected onto the $x$--$y$ plane for visualization and (b) shown in 3D. (c) The Jaccard index curves of the three results at different depths. (d) The RMSE of the three results at different depths.}
\label{fig:loca}
\end{figure}

To further validate the feasibility of our method, we demonstrated a 3D localization experiment of 1 $\upmu$l fluorescent beads ($1$ $\upmu$m diameter, type F10002) uniformly immersed into 10 ml agarose. An epifluorescence microscope mentioned above was used to collect the wide-field fluorescence image, as shown in Fig.~\ref{fig:loca}(a). We still utilized the trained CNNs for localization as aforementioned with the single captured stack. Compared with the simulated point spread function, the experimental calibrated focal stack of the particular sample takes the system aberration and noise condition into consideration and makes the learning algorithm adaptive to different samples with better performance and a more convenient operation instead of complicated parameter selections. The trained networks were then applied to the captured fluorescence image for localization in a volume of $200\times200\times100$ $\upmu$m$^3$, as shown in Fig.~\ref{fig:loca}(a) and Fig.~\ref{fig:loca}(b). The ground truth positions of the beads, marked by red spots, were retrieved from a deconvolved wide-field focal stack of the same agarose volume (see Fig.~3 in supplementary material).

The localization results of our method and MLE method are marked by green squares and yellow triangles, respectively. The Jaccard index and RMSE of the results at different axial planes, conducted by our method and other kernel fitting methods such as MLE and least squares (LS) \cite{mortensen2010optimized}, are shown in Fig.~\ref{fig:loca}(c) and Fig.~\ref{fig:loca}(d), respectively. Although all of the algorithms gradually lose the recognition accuracy and localization precision with the increase of the depth, our method has a superior performance in both aspects, which can be obviously observed in Fig.~\ref{fig:loca}(b).

\subsection{\emph{In Vivo} 3D Cell Tracking in Zebrafish}

\begin{figure}[t]
\centering
\includegraphics[width=.75\linewidth]{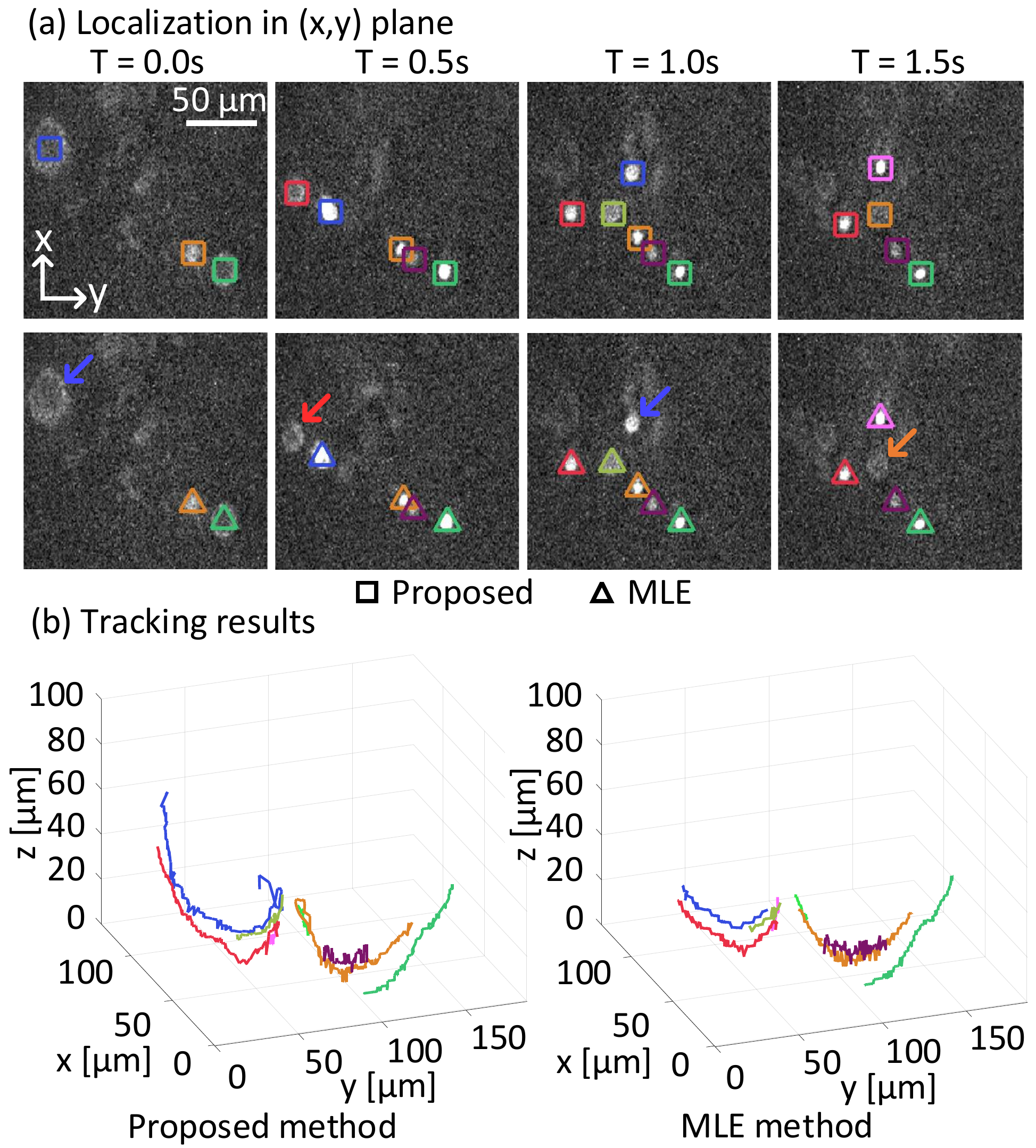}
\caption{Tracking blood cells at 100 fps of a one-day-old live zebrafish restrained in agarose. (a) Captured wide-field fluorescence images of the ROI at different time stamps and the corresponding localization results reconstructed by our method and MLE method. (b) 3D tracking results of the blood cells reconstructed by our method and MLE method.}
\label{fig:fish2}
\end{figure}

Benefiting from the capability of localizing multiple objects in 3D from a single wide-field fluorescence image, we can realize fast 3D cell tracking at any normal commercial fluorescence microscope without any hardware modification. For verification, we conducted several experiments on zebrafish to track the 3D movement of multiple blood cells. One-day-old Tg(gata1:GFP) zebrafishes expressing GFP in the blood cell were used in our experiments. The fishes were live, awake, and immobilized in $1\%$ low-temperature melting agarose. We used the same experimental setup as that of the beads experiment described previously. It needs to be noticed that the size of the tagged fluorophore in a blood cell was about 5 $\upmu$m, leading to a different diffraction pattern from the bead. To improve the localization accuracy, we recorded a new focal stack of a zebrafish blood cell to synthesize sufficient samples for the network training (see Fig.~4 in supplementary material). When capturing the focal stack of the zebrafish blood cell, the zebrafish was anaesthetized and euthanized with $0.2\%$ tricaine. This step can be viewed as a calibration process, which is conducted only once for one specific sample.

Then we took a $1.5$-second video at $100$ fps to record the motion of the blood cells close to the heart. Due to the heart beating, the speed of the blood cell is around $75$ $\upmu$m/s, which is very fast and hard to be tracked in 3D with conventional methods. The images of one region of interest (ROI) at different time stamps are shown in Fig.~\ref{fig:fish2}(a). The first row shows the localization results of our method, marked by squares, while the second row shows the results of MLE method, marked by triangles for comparison. Different colors correspond to different cells. After acquiring the localization result of each frame, we used a Kalman filter-based tracking algorithm to get the temporal traces of different cells in 3D, as illustrated in Fig.~\ref{fig:fish2}(b). Compared with the MLE method, our 3D tracking has a much larger axial range with a smoother temporal trace, taking advantage of its higher localization accuracy. Another experimental result with more blood cells at lower speed is shown in Fig.~5 in supplementary material. For a quantitative analysis of the tracking performance, we conducted a numerical simulation to show the comparison of the tracking results with respect to different moving patterns, as shown in Fig.~6 in supplementary material.

\section{Discussion}

\begin{figure}[t]
\centering
\includegraphics[width=.75\linewidth]{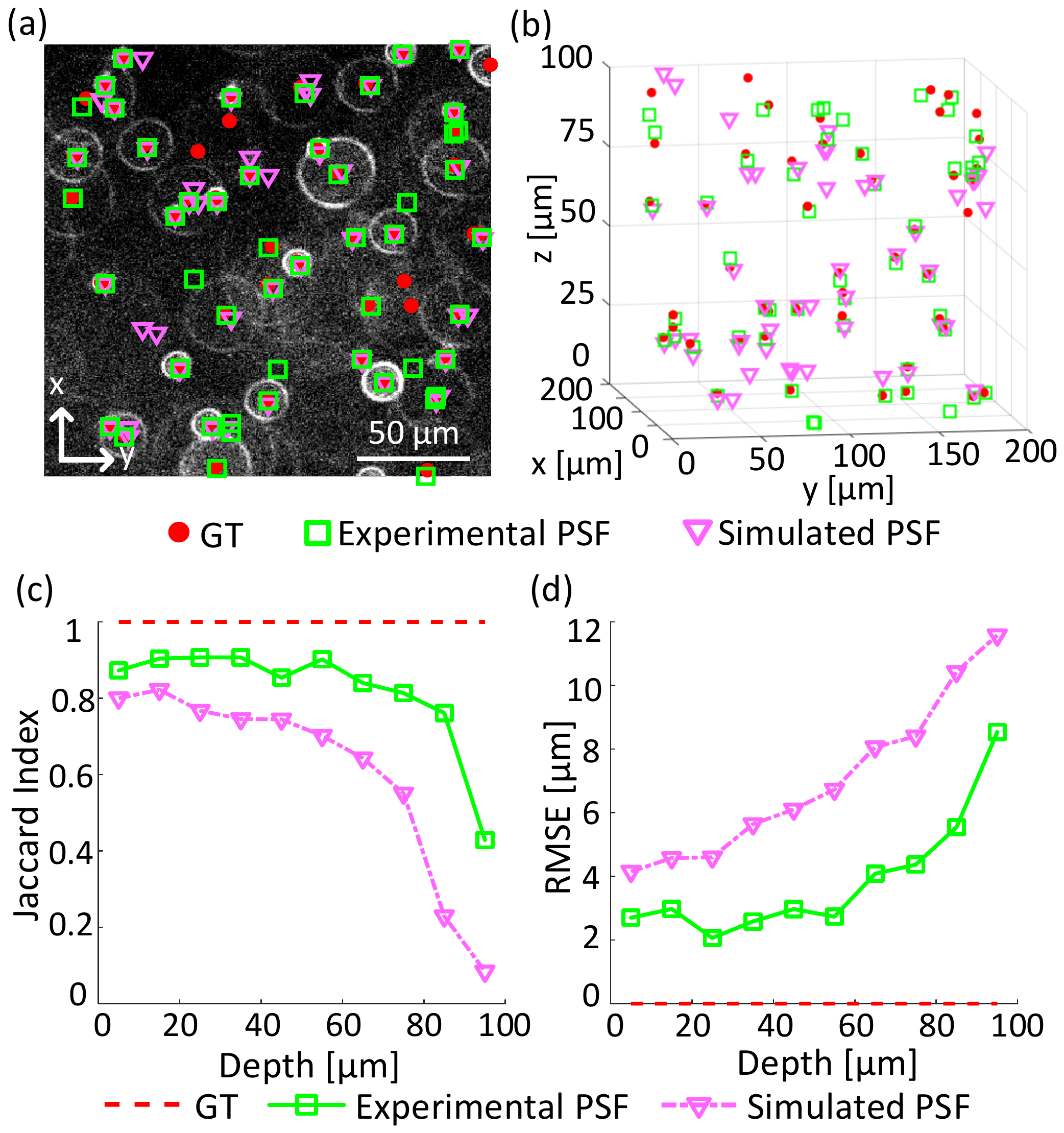}
\caption{Single-shot wide-field fluorescence image and 3D localization of randomly placed fluorescent beads. (a) and (b) Collected wide-field fluorescence image and comparison of the localization results of the trained CNNs with the training samples generated by experimental PSF (our method) and simulated PSF. (a) Estimated source positions are projected onto the $x$--$y$ plane for visualization and (b) shown in 3D. (c) The Jaccard index curves of the two results at different depths. (d) The RMSE of the two results at different depths.}
\label{fig:loca2}
\end{figure}

With the increasing attentions of deep learning, numerous learning based algorithms have been proposed in optics to show its unique advantages over traditional methods. In this work, we present a new deep learning-based framework for 3D localization of multiple objects from a single wide-field fluorescence image. By converting the 3D localization problem into a cascaded classification problem, we make full use of the superb ability of deep convolutional neural network in supervised learning \cite{lecun2015deep}. Compared to traditional maximum-likelihood algorithms, our method is easier to calibrate, orders of magnitude faster (once trained), and has better performance in cases of poor SNR and large sample density, which can hardly be attacked by the traditional localization methods due to their congenital limitations.

The large amounts of training data for our framework is obtained from the simulation of the incoherent superposition of multiple objects with the prior knowledge of the z-stack of a single object. While the z-stack of the object can be synthesized by the simulated point spread function (PSF) and the shape of the object, we choose an experimental z-stack of a single object for training data synthesis due to the diverse imaging environments in different experiments, such as the optical aberration, medium induced refractive index mismatch, and noise condition. To show the differences in detail, we trained another CNNs with the simulated PSF for the localization of the fluorescent beads in Fig.~\ref{fig:loca} and compared the results obtained by the CNNs with the experimental z-stack of a single fluorescent bead, as illustrated in Fig.~\ref{fig:loca2}. Both the recognition and localization accuracy of the networks with totally simulated data are much lower at different axial positions in terms of Jaccard index and RMSE.

The intrinsic limitation of most localization methods comes from the total photons emitted by the tracking object such as a single molecule or fluorescent bead, which usually has an upper limit set by the photochemical stability and the maximum emission rate of the molecule \cite{von2017three}. For 3D localization, we retrieve the 3D position of the object from a single 2D image by using all of the defocus information. In this case, with the increase of the defocus distance, the point spread function becomes larger and the signal photon density gets lower, while the shot noise from the background fluorescence and the readout and dark noise from the sensor remain the same. Thus different axial planes have different SNR conditions for localization, which results in the limited localization range of conventional kernel fitting methods. We take this SNR difference among different axial planes into account by generating our training data with the experimental focal stack for calibrating the noise condition. That's the reason for a much larger axial localization range of our method compared with MLE method, which also facilitates the 3D cell localization from a single 2D wide-field fluorescence image.

Indeed, we still suffer from a performance degradation at very far from the focal plane due to the extremely low photon density with large defocus. Such limitation can be definitely overcome by introducing the PSF engineering methods \cite{shechtman2016multicolour, wang2017single}. With a specific phase modulation at the Fourier plane, the PSF of the microscope can be modulated to be depth-variant and convergent in a much larger axial range, compared with normal wide-field fluorescence microscope. In the future, we will demonstrate more applications in biology such as high-speed large-scale tumor metastasis monitoring in zebrafish and egg development of different species. In addition, other imaging modalities, such as bright-field imaging and phase contrast imaging, can exploit this framework to facilitate even broader clinical applications like monitoring the 3D traces of multiple freely moving sperms in a wide field.

\section{Conclusion}

In summary, we propose a novel 3D localization framework for multiple objects with two cascaded deep convolutional neural networks. The proposed method converts the challenging 3D localization of multiple objects with great crosstalk from a ill-posed model-based fitting problem to a multi-label classification problem, where deep learning technique has a great advantage over other algorithms. Our method achieves higher recognition accuracy and lower localization error than the state-of-the-art kernel fitting methods, which is validated by both simulation and experimental results on 3D localization of fluorescent beads in agarose. With our technique, fast 3D cell tracking can be conducted by any commercial fluorescence microscope without hardware modifications. \emph{In vivo} 3D tracking of multiple blood cells in zebrafish at 100 fps demonstrates its feasibility. Combining with other PSF engineering technologies, the limited localization range of our method can be further enhanced. In addition to more biological applications, such as multiple-molecule tracking and the monitoring of tumor metastasis, we anticipate that our method can guide more applications of deep learning in the optics field.





\section*{Funding Information}
National Natural Science Foundation of China (NSFC) (61327902, 61722209).

\section*{Acknowledgments}
The authors thank Zheng Jiang for the sample preparation.

\section*{Supplemental Documents}

\bigskip \noindent See supplementary material for supporting content.

\bigskip \noindent $^\dag$These authors contributed equally to this work.

\end{document}